\documentclass[conference]{IEEEtran}
\IEEEoverridecommandlockouts
\usepackage{cite}
\usepackage{amsmath,amssymb,amsfonts}
\usepackage{algorithmic}
\usepackage{graphicx}
\usepackage{textcomp}
\usepackage{xcolor}
\usepackage{booktabs}
\usepackage{lipsum}
\usepackage[normalem]{ulem}

\def\BibTeX{{\rm B\kern-.05em{\sc i\kern-.025em b}\kern-.08em
    T\kern-.1667em\lower.7ex\hbox{E}\kern-.125emX}}

\usepackage{comment}
    
\begin{document}

\title{Early feasibility of an embedded bi-directional brain-computer interface for ambulation\\
\thanks{This work was funded by the National Science Foundation (Award No. 1446908, 1646275)}
}


\author{
\IEEEauthorblockN{Jeffrey Lim\IEEEauthorrefmark{1}, Po T. Wang\IEEEauthorrefmark{1}, Wonjoon Sohn\IEEEauthorrefmark{1}\IEEEauthorrefmark{6}, Claudia Serrano-Amenos\IEEEauthorrefmark{1}, Mina Ibrahim\IEEEauthorrefmark{1}, Derrick Lin\IEEEauthorrefmark{2},\\ Shravan Thaploo\IEEEauthorrefmark{2}, Susan J. Shaw\IEEEauthorrefmark{3}, Michelle Armacost\IEEEauthorrefmark{3}, Hui Gong\IEEEauthorrefmark{3}, Brian Lee\IEEEauthorrefmark{3}\IEEEauthorrefmark{4}, Darrin Lee\IEEEauthorrefmark{3}\IEEEauthorrefmark{4}, Richard A. Andersen\IEEEauthorrefmark{5},\\ Payam Heydari\IEEEauthorrefmark{1}, Charles Y. Liu\IEEEauthorrefmark{3}\IEEEauthorrefmark{4}, Zoran Nenadic\IEEEauthorrefmark{1}, and An H. Do\IEEEauthorrefmark{2}}
\IEEEauthorblockA{\IEEEauthorrefmark{1}University of California -- Irvine (UCI), Irvine, CA, USA\\ 
Email: \{limj4, znenadic\}@uci.edu
}
\IEEEauthorblockA{\IEEEauthorrefmark{6}Work done at UCI, currently at Abbot Laboratories, Plano, TX, USA}
\IEEEauthorblockA{\IEEEauthorrefmark{2}UCI School of Medicine, Irvine, CA, USA\\ 
Email: and@uci.edu
}
\IEEEauthorblockA{\IEEEauthorrefmark{3}Rancho Los Amigos National Rehabilitation Center, Downey, CA, USA
}
\IEEEauthorblockA{\IEEEauthorrefmark{4}Keck School of Medicine of University of Southern California (USC), USC Neurorestoration Center, Los Angeles, CA, USA}
\IEEEauthorblockA{\IEEEauthorrefmark{5}California Institute of Technology, Pasadena, CA, USA
}

} 

\maketitle

\begin{abstract}
Current treatments for paraplegia induced by spinal cord injury (SCI) are often limited by the severity of the injury. The accompanying loss of sensory and motor functions often results in reliance on wheelchairs, which in turn causes reduced quality of life and increased risk of co-morbidities. While brain-computer interfaces (BCIs) for ambulation have shown promise in restoring or replacing lower extremity motor functions, none so far have simultaneously implemented sensory feedback functions. Additionally, many existing BCIs for ambulation rely on bulky external hardware that make them ill-suited for non-research settings. Here, we present an embedded bi-directional BCI (BDBCI), that restores motor function by enabling neural control over a robotic gait exoskeleton (RGE) and delivers sensory feedback via direct cortical electrical stimulation (DCES) in response to RGE leg swing. A first demonstration with this system was performed with a single subject implanted with electrocorticography electrodes, achieving an average lag-optimized cross-correlation of 0.80$\pm$0.08 between cues and decoded states over 5 runs.  
\end{abstract}


\section{Introduction}
Individuals with paraplegia after spinal cord injury (SCI) typically experience a loss of motor and sensory function in the lower extremities. While therapies exist to compensate for these neurological deficits, the severity of the SCI may limit their effectiveness. This often leads to a reliance on wheelchairs, which can cause co-morbidities (heart disease, osteoporosis, pressure ulcers~\cite{johnson_cost_1996}) that result in reduced productivity as well as lower quality of life. As such, people with paraplegia place ambulation restoration as a top rehabilitation priority~\cite{anderson_targeting_2004}. Brain-computer interfaces (BCIs) promise to fulfill this unmet need by re-establishing brain control over paralyzed extremities. An example of such a BCI is the non-invasive system described in~\cite{king_brain-computer_2014}, which enabled a subject with paraplegia wearing an electroencephalography (EEG) cap to control functional electrical stimulators (FESs) attached to their leg muscles. Recently, Benabid et al.~\cite{benabid_exoskeleton_2019} developed an invasive BCI system utilizing the superior spectral/temporal properties of electrocorticography (ECoG) signals~\cite{mccrimmon_electrocorticographic_2018} to allow a subject with tetrapelgia to walk by controlling a robotic gait exoskeleton (RGE). 

Despite these advances, a key shortcoming for existing BCIs for ambulation is that they lack somatic sensory feedback and instead rely primarily on visual feedback for closed-loop control. Non-invasive BCI in particular have no readily available method for restoring somatic sensory feedback. In contrast, invasive BCI can potentially incorporate artificial somatic sensation via direct cortical electrical stimulation (DCES) of sensory areas~\cite{hiremath_human_2017,lee_engineering_2018}. These ``bi-directional" BCI (BDBCI) restore both motor and sensory pathways, thereby achieving more biomimetic sensorimotor restoration and potentially enhancing overall BCI control~\cite{flesher_intracortical_2017}. Furthermore, in order to enable long-term, non-research usage for BDBCIs, future designs will likely need to be fully-implantable. This necessitates an embedded systems approach, instead of relying on bulky external hardware that many existing lower extremity BCI systems depend on to achieve real-time decoding~\cite{king_brain-computer_2014,benabid_exoskeleton_2019}.

To address these issues, we have developed an ECoG-based, embedded BDBCI system combining real-time motor decoding~\cite{wang_benchtop_2019} with DCES for artificial sensory feedback~\cite{sohn_benchtop_2022}. While aspects of this system have been tested on the benchtop~\cite{sohn_benchtop_2022}, the integration of decoding and stimulation functions alongside RGE control have yet to be demonstrated. In this work, we describe the performance of a BDBCI-RGE system in a human subject implanted with ECoG electrodes. This includes the ability to decode leg motor activity in real time from primary motor cortex (M1) ECoG signals, actuate RGE stepping according to the decoded intent, and deliver DCES to the primary sensory cortex (S1) to elicit artificial leg sensory feedback in response to RGE leg swing. This study represents a first-of-a-kind demonstration of a BDBCI system for ambulation.

\section{Methods}\label{sec:methods}
\subsection{System Overview} \label{sec:system}
\begin{figure}
    \centering
    \includegraphics[width=\linewidth]{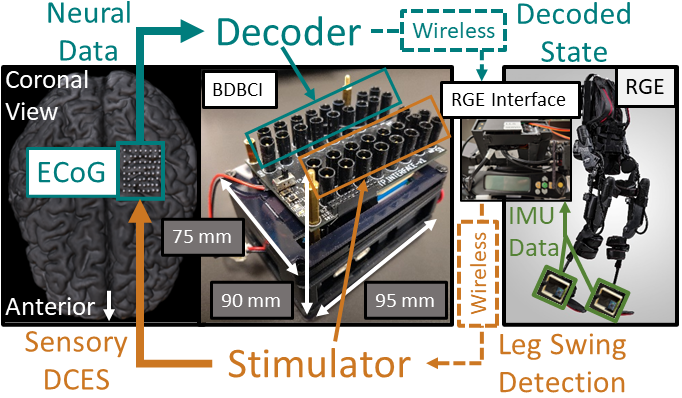}
    \caption{BDBCI-RGE system overview. Detailed explanation in Section~\ref{sec:system}.}
    \label{fig:eksobdbci_overview}
\end{figure}

The BDBCI-RGE system (Fig.~\ref{fig:eksobdbci_overview}) consists of the BDBCI, an RGE (Ekso-GT, Ekso Bionics, Richmond, CA, USA), and an RGE interface. Data from ECoG electrodes over M1 were acquired by the BDBCI, which then performed a binary classification on those data to determine the decoded state (``Move" or ``Idle"). These states were then wirelessly transmitted to the RGE interface, which would trigger the RGE to step if the received state was ``Move". Inertial measurement units (IMUs) attached to the RGE recorded leg kinematic data, which were used by the RGE interface to instruct the BDBCI to stimulate S1 in response to leg swing to elicit sensory feedback.

\subsection{Hardware Design} \label{sec:hardware}
 The BDBCI was implemented on a custom designed PCB (See~\cite{sohn_benchtop_2022}). Briefly, onboard microcontrollers (48 MHz, Microchip, Chandler, AZ) executed all system functions, namely acquisition, wireless communication, decoder model generation, online decoding, and DCES. Neural data was acquired from up to 16 ECoG electrodes (16 recording channels with a reference electrode) at 500 Hz using an amplifier array integrated circuit (Intan Technologies, Santa Monica, CA). Wireless transceivers (RFM98, HOPE Microelectronics, Xili, ShenZhen, China) enabled communication with a base station computer to log experimental data. A custom GUI on the base station computer was used to initiate or abort online decoding.

The custom GUI was also used to set stimulation parameters. Here, operators could select the bipolar stimulation channel (any pair of electrodes) and waveform properties (pulse train frequency, anodic/cathodic pulse width, current amplitude, pulse train duration). The BDBCI stimulator could then output biphasic square-pulse trains with the chosen parameters. After parameters were set, the stimulation could then be triggered either manually from the GUI (for cortical mapping, see Section~\ref{sec:stim}) or in response to leg swings detected by the RGE interface.

The RGE interface was comprised of a microcontroller and radio transceiver (both same as above) that enabled wireless communication with the BDBCI. A servomotor connected to the RGE interface depressed the ``Step" button on the RGE's controller upon receipt of a decoded "Move" state from the BDBCI. The RGE interface was also connected to two 6-axis IMUs (TDK Corporation, Tokyo, Japan), which were mounted on the ``ankles" of the RGE legs. The angular velocity derived from the IMU data was used to detect the initiation and completion of each RGE step. Based on this information, the RGE interface wirelessly instructed the BDBCI to deliver DCES during leg swing to elicit sensory feedback.

\subsection{ECoG Procedures and Online Decoding Model Generation} \label{sec:decode}
Subjects were recruited from a population of patients undergoing ECoG implantation for epilepsy surgery evaluation with electrode coverage of M1 and S1 areas with expected leg representation. Electrodes with power modulation in the $\mu$-$\beta$ (8--25 Hz) and/or high-$\gamma$ (80--160 Hz) bands in response to leg motor movement were plugged into the BDBCI. Specifically, we sought electrodes exhibiting $\mu$-$\beta$ desynchronization and/or high-$\gamma$ synchronization during leg movement~\cite{mccrimmon_electrocorticographic_2018}. To identify such electrodes, the subject was asked to perform alternating periods of idling and leg movement while ECoG data from all electrodes were recorded using an ICU neural monitoring system (Natus\textsuperscript{\textregistered} Quantum$^\textnormal{TM}$, Natus Medical Incorporated, Pleasanton, CA). These data were then processed in MATLAB and visually inspected to identify the subset of electrodes exhibiting $\mu$-$\beta$ and/or high-$\gamma$ modulation.

Prior to real-time BDBCI operation, an online decoding model must be generated by first collecting ECoG signals from the previously identified electrodes during idling and leg movement behavior and then calculating the corresponding feature extraction matrices~\cite{wang_benchtop_2019}. These matrices were used in real time to obtain features on which a binary state classifier can be applied to obtain a decoded state. A training data collection protocol was implemented on the BDBCI, which was initiated via the GUI. During this protocol, the subject was seated in the ICU bed (configured into an upright, seated position) and was prompted by a screen displaying ``Idle" or ``"Move" cues that alternated every 10 seconds for a total of 80 seconds. The subject was instructed to hold still and relax when ``Idle" was displayed, and to perform a seated ``marching" motion with both legs when ``Move" was displayed. The BDBCI acquired and saved the ECoG signals 
in onboard memory for online decoding model generation.

To calculate the feature extraction matrices from the training data, the data was first common-average referenced. Each "Move" and "Idle" trial was then subdivided into non-overlapping 750 ms windows. For each window and channel, the average power in the $\mu$-$\beta$ and the high-$\gamma$ band was calculated. These powers were concatenated across the two bands and channels, nominally resulting in 2$\times$16 data. These 32-dimensional data were then processed using a class-wise principal component analysis (cPCA)~\cite{nenadic_information_2007}.
This produced a cPCA matrix for each class (``Idle" or ``Move") that reduced the dimension of the data. Linear discriminant analysis (LDA) was then applied to enhance the separability between the two classes~\cite{fisher_use_1936,hart_pattern_2000}, as well as to further reduce the dimension of the data to a one-dimensional feature. The combined cPCA-LDA feature extraction matrix for each class was then saved onboard for use in real-time decoding.

\subsection{Real-Time Decoding and Online Model Validation} \label{sec:online}
When performing real-time decoding, the BDBCI acquires data in 250 ms common-average referenced windows. The decoder then calculates the $\mu$-$\beta$ and high-$\gamma$ powers for each channel and averages these over the four most recent windows. The saved cPCA-LDA matrices then reduce the dimension of the data to a one-dimensional feature for each class. A binary Bayesian classifier~\cite{hart_pattern_2000} is then applied to this feature to select the decoded state for the most recent data window.

We used short real-time decoding experiments to validate the online decoding model. Specifically, we instructed the subject to follow ``Idle" and ``Move" cues that alternated every 10 seconds for a total of 80 seconds while the BDBCI decoded the ECoG data in real time. The resulting decoded states and corresponding cues were simultaneously logged by the base station computer. The performance of online decoding models were then evaluated by calculating the lag-optimized cross-correlation between the decoded state and the cues.

\subsection{Sensory Stimulation Mapping Procedure}\label{sec:stim}
Cortical stimulation mapping was used to find stimulation parameters with sensory responses. Adjacent pairs of ECoG electrodes in S1 were sequentially connected to the BDBCI to form bipolar stimulation channels. For each channel, stimulation was delivered via the GUI using 1 s pulse trains (250 $\mu$s/phase). Pulse train frequency and current amplitude were varied from 50 to 300 Hz and 2 to 10 mA, respectively, until a response was elicited or the maximum value was reached. For each parameter set, the subject was asked for a verbal description of any elicited sensations. We documented those parameter sets eliciting sensation in the leg contralateral to ECoG implantation. Ultimately, we designated one parameter set for BDBCI operation.

\subsection{Online BDBCI-RGE Validation} \label{sec:exp_design}
\begin{figure}
    \centering
    \includegraphics[width=0.90\linewidth]{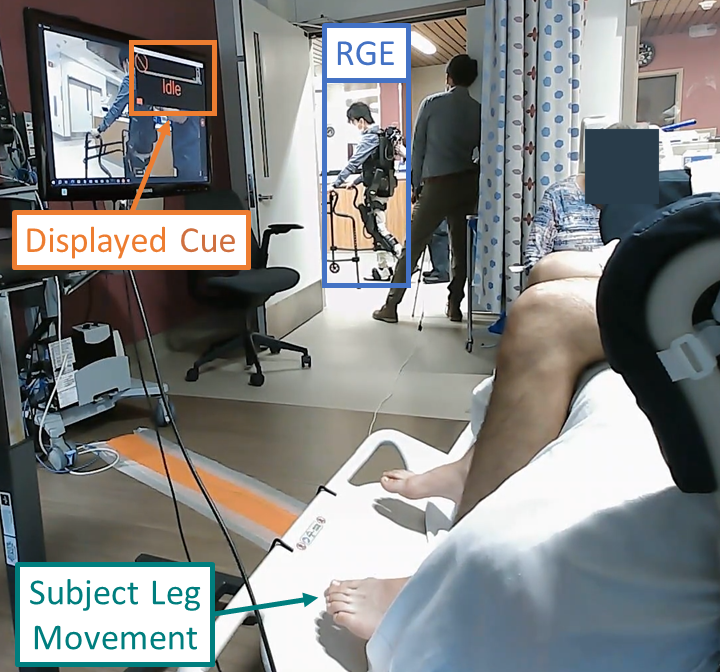}
    \caption{BDBCI-RGE walking task. Subject is connected to BDBCI (off-screen) and remains in the ICU bed while following displayed cues to wirelessly control the RGE in real time.}
    \label{fig:RGE_walking_task}
\end{figure}

Once the online decoding model was generated (Section~\ref{sec:decode}) and stimulation parameters were chosen (Section~\ref{sec:stim}), the subject proceeded to the BDBCI-RGE walking task. Here, an experimenter was placed in the RGE in the subject's view, while the BDBCI decoded walking states from the subject in real time (similar to Section~\ref{sec:online}). Note that in the BDBCI-RGE walking task, the BDBCI suspends acquisition during RGE leg swing to avoid electrical interference from stimulation. The subject remained in the ICU bed (see Fig.~\ref{fig:RGE_walking_task}) while following ``Idle" or ``Move" cues that alternated every $\sim$25 s for a total of $\sim$125~s. The decoded states were wirelessly transmitted to the RGE for real-time control. The subject also received DCES with the predetermined parameters (Section~\ref{sec:stim}) in response to RGE leg swing (Section~\ref{sec:hardware}). This BDBCI-RGE walking task was repeated for 5 runs. Using the logged decoded state and cue data, the online BDBCI performance was evaluated for each run as described in Section~\ref{sec:online}.

\section{Results}
\begin{figure}
    \centering
    \includegraphics[width=\linewidth]{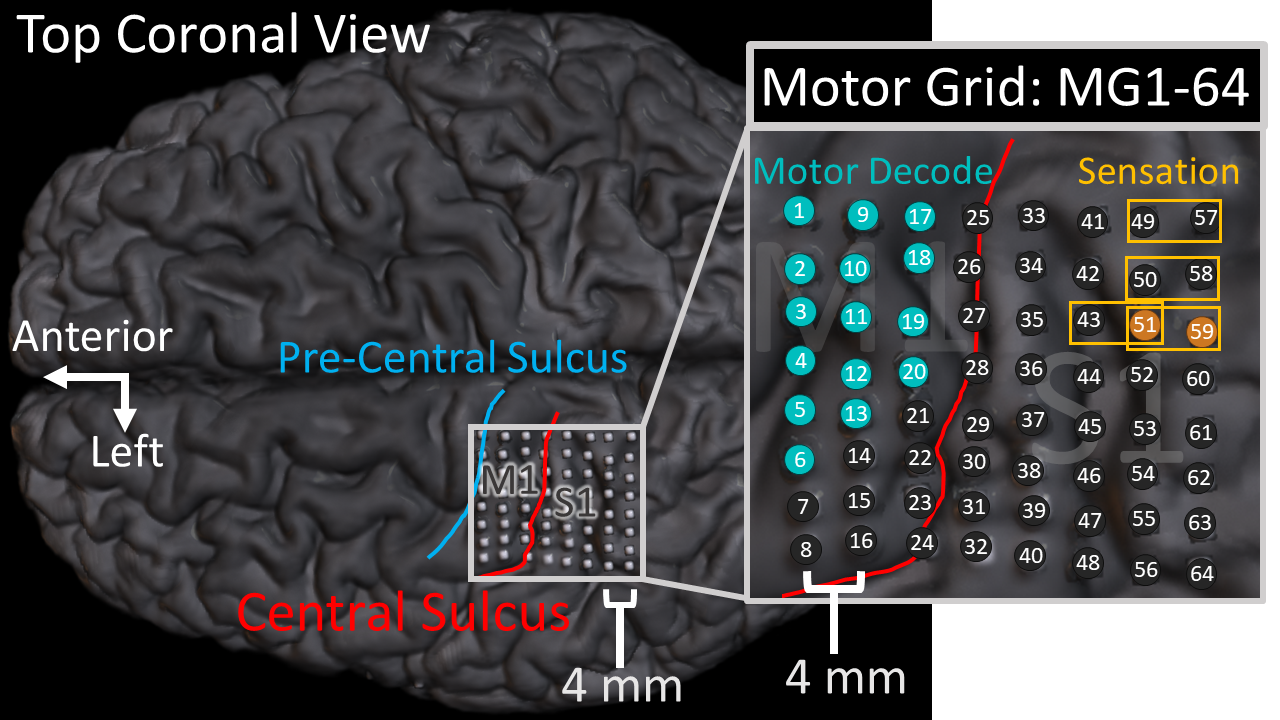}
    \caption{Coregistration of ECoG electrodes and subject's brain segmented from post-implant CT image and pre-implant MR image, respectively. Approximate locations of the pre-central sulcus (blue) and central sulcus (red) delineate the M1 and S1 cortices. (Teal): electrodes plugged to the BDBCI for motor decoding. (Yellow boxes): Stimulation channels with right leg sensory responses. (Orange): stimulation channel used for BDBCI-RGE task. (Dark grey): unplugged electrodes.}
    \label{fig:coreg}
\end{figure}
\subsection{ECoG Subject Details}
This study was approved by the IRB of the University of California, Irvine and the Rancho Los Amigos National Rehabilitation Center. One subject (age 22, M), undergoing epilpesy surgical evaluation with ECoG implanted over the left M1 and S1 areas with expected leg representation (Fig.~\ref{fig:coreg}), provided written informed consent to participate in the study. Following the protocol described in Section~\ref{sec:decode}, we chose 15 electrodes in M1 for motor decoding (see Fig.~\ref{fig:coreg}). MG63 and MG64 were used as reference and ground, respectively.

\subsection{Online Decoding Model Generation}\label{sec:result_decode}
\begin{figure}
    \centering
    \includegraphics[width=\linewidth]{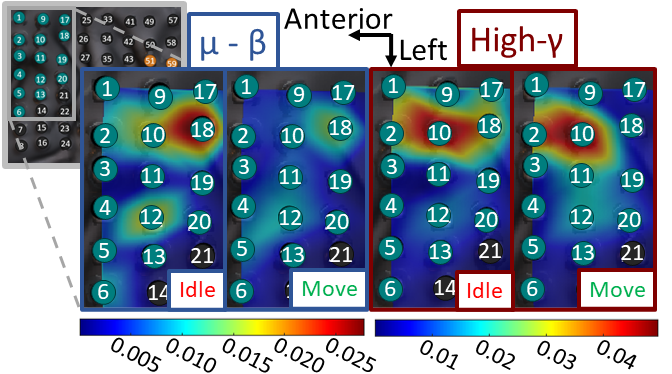}
    \caption{Spatial weights for cPCA-LDA feature extraction matrices color-coded and mapped to ECoG electrode positions. Only electrodes plugged to BDBCI for motor decoding (teal) are given values, unused electrodes (grey) are assigned a value of zero. Values between electrodes are linearly interpolated. Electrodes with larger weight values contain features that are more relevant to the online decoding model.}
    \label{fig:fmats}
\end{figure}

\begin{figure}
    \centering
    \includegraphics[width=\linewidth]{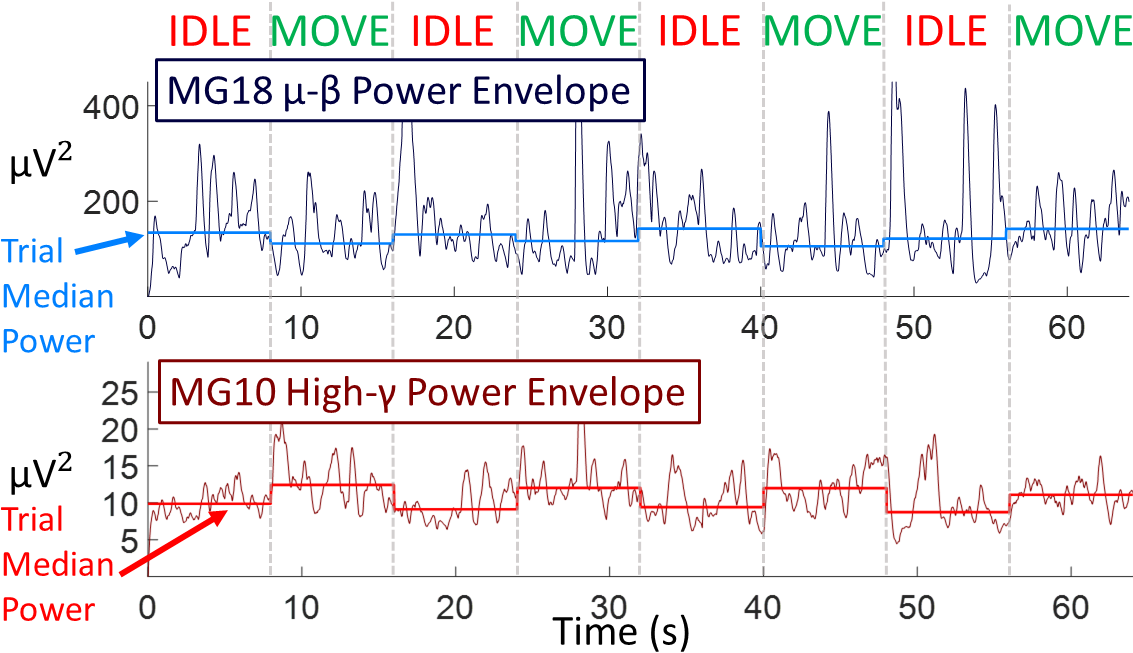}
    \caption{Representative 1-Hz power envelopes exhibiting in-band modulation from decoder training data. Weight values (Fig. \ref{fig:fmats}) indicate MG18 and MG10 contained features salient for decoding. (Top): $\mu$-$\beta$ modulation on MG18: Trial median power generally increases during ``Idle" relative to ``Move". (Bottom): high-$\gamma$ modulation on MG10: Trial median power increases during ``Move" relative to ``Idle".}
    \label{fig:ecogmodulation}
\end{figure}

 An online decoding model was built as described in Section~\ref{sec:decode}. Fig.~\ref{fig:fmats} shows the spatial weights of the resulting combined cPCA-LDA feature extraction matrices, illustrating electrodes containing salient features for decoding. More specifically, MG18 and MG12 contained salient features in the $\mu$-$\beta$ band for both ``Idle" and ``Move" states. Similarly, MG10 contained salient features in the high-$\gamma$ band for both states. Naturally, these electrodes exhibited modulation in the corresponding bands (Fig. \ref{fig:ecogmodulation}) and formed the basis for neural walking control. An online validation (without RGE or stimulation) yielded a lag-optimized cross-correlation between decoded states and cues of 0.81 (decoded state lag of 1 s).
 
\subsection{Stimulation Mapping Results}
Stimulation mapping was performed over the electrodes in the first three rows of the medial-posterior quadrant of the ECoG grid, as these electrodes were expected to be closest to the leg sensory area. Stimulation parameters that elicited right leg sensory responses and the subject's verbal descriptions are reported in Table~\ref{tab:stim_map}. Though some motor responses (involuntary movement) were also elicited, channels with motor responses at any waveform parameters were excluded. Ultimately, stimulation channel MG51-59 (6.2 mA current amplitude, 300 Hz pulse train frequency, 250 $\mu$s anodic/cathodic pulse width) was chosen for the BDBCI-RGE walking task. 

\begin{table}
    \centering
    \caption{Elicited sensory responses}
    \begin{tabular}[width = \linewidth]{p{1.25cm} p{1cm} p{1.5cm} p{1.5cm} p{1.25cm}}
    \toprule
         Stimulation Channel & Amplitude (mA) & Pulse Train Frequency (Hz) & Sensation Location (Right Leg) & Reported Sensation \\
    \midrule
         MG49-57 & 9.9 & 100 & Big Toe & Tingling \\
         MG50-58 & 7.4 & 300 & Heel & Tingling \\
         MG51-59 & 6.2 & 300 & Heel & Tingling \\
         MG43-51 & 6.9 & 200 & Heel & Tingling \\
    \bottomrule
    \end{tabular}
    \label{tab:stim_map}
\end{table}

\subsection{BDBCI-RGE Walking Task Results}
\begin{figure}
        \centering
        \vspace{-0.4 cm}
        \includegraphics[width=\linewidth]{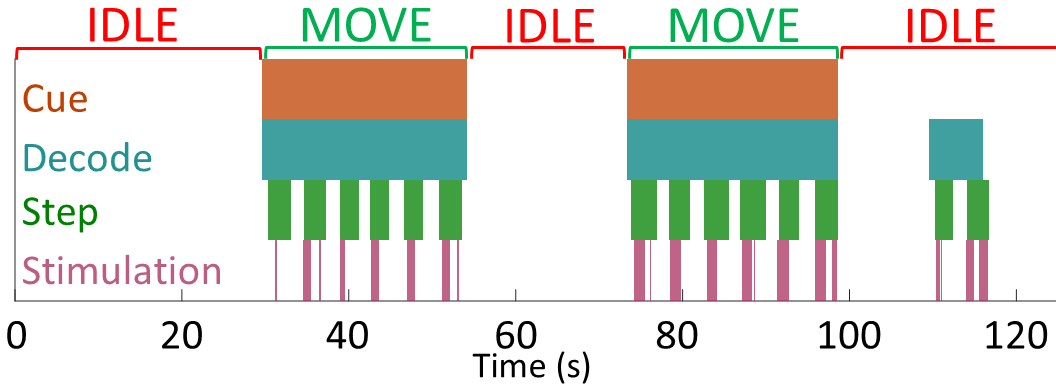}
        \caption{Online decoding for Run \#2 for BDBCI-RGE walking task. Colored blocks for cue and decode traces indicate ``Move" periods. Colored blocks for step trace indicate duration of RGE step as extracted from video. Colored blocks for stimulation trace indicate stimulation is being delivered. Cues and decoded state are generally well correlated for this run, except for a period of decoded move causing two steps in the last ``Idle" trial.}
        \label{fig:cuedecodestepstim}
    \end{figure}

    The performance results for the five online decoding runs of the BDBCI-RGE walking task are summarized in Table~\ref{tab:BDBCI_corr_table}. An average cross-correlation between cues and decoded state of 0.80 $\pm$ 0.08 was achieved, which is comparable to the decoding performance without the RGE or stimulation (Section~\ref{sec:result_decode}). Fig.~\ref{fig:cuedecodestepstim} visualizes one of the online decoding runs, illustrating the concurrence between cues and decoded state, as well as between step and stimulation events. Imprecise synchronization between computer and BDBCI clocks led to a few reports of stimulation onset preceding step initiation, however the two events were still largely one-to-one. 

\begin{table}
    \centering
    \vspace{-0.4 cm}
    \caption{BDBCI-RGE Online Decoding Performance}
    \label{tab:BDBCI_corr_table}
    \begin{tabular}{ccccccc}
        \toprule
         Run \# & 1 & 2 & 3 & 4 & 5  \\
         \midrule
         Max X Corr.& 0.88 & 0.90 & 0.76 & 0.72 & 0.76  \\ 
         Lag (s) & 0 & 0 & 0 & 8 & 0\\
         \bottomrule
    \end{tabular}
\end{table}
    
\section{Discussion}
This study represents a first demonstration of the feasibility of BDBCI for human ambulation. Notably, we implemented DCES for artificial sensory feedback alongside real-time neuromotor BDBCI control, which the subject quickly established within 3 days. While examples of BDBCI systems already exist for the upper extremities~\cite{flesher_intracortical_2017}, our system marks the first BDBCI for lower extremity applications.

Our system is also the first embedded system BDBCI for ambulation. An embedded system approach is critical for the implementation of BDBCI in a fully-implantable form factor, which will facilitate long-term usage of BDBCIs outside of a research setting. Whereas most invasive BCIs are dependent on large external hardware, our system exclusively utilizes onboard hardware to perform real-time decoding and DCES functions. Note that the base station computer is not necessary for online decoding, as it primarily serves to enable experimental procedures. Benabid et. al~\cite{benabid_exoskeleton_2019} and Lorach et. al~\cite{lorach_walking_2023} employed an implantable embedded system~\cite{mestais2014wimagine} to acquire and transmit neural information for their lower-extremity BCIs, but relied on body-mounted laptops/computers to perform decoding computations. Additionally, this implant was designed to fit to the lateral convexity of the brain, which may prove to be a limitation as leg/gait representation is anatomically located in the inter-hemispheric space~\cite{mccrimmon_electrocorticographic_2018}. Though our study also utilized ECoG grids implanted on the cortical convexity, we intend to use inter-hemispheric grids in actual SCI patients in the future. Since leg representation in the inter-hemispheric space is richer, this could potentially optimize the performance of leg motor decoding (i.e. to surpass the performance in EEG-BCI~\cite{king_feasibility_2015,do_brain-computer_2013}) and elicit more naturalistic leg sensations.

Future work will seek to test our BDBCI-RGE system with SCI subjects with paraplegia. Ideally, SCI subjects will be placed in the exoskeleton for these tests. We will also pursue system optimizations to improve data logging and online operation. For example, decoding delays due to interleaving of acquisition and stimulation may have contributed to suboptimal decoding performance. This could be avoided by using artifact-suppressing hardware~\cite{pu_cmos_2021} and digital algorithms~\cite{lim_artifact_2022} to enable full-duplex bi-directional operation.

\bibliographystyle{ieeetr}

\end{document}